\documentclass[reprint,
superscriptaddress,
groupedaddress,
unsortedaddress,
runinaddress,
frontmatterverbose, 
nofootinbib,
nobibnotes,
bibnotes,
amsmath,amssymb,
aps,
pra,
floatfix,
]{revtex4-2}
\usepackage[dvipsnames]{xcolor}
\usepackage{graphicx,
subfig,
lipsum
}
\usepackage{xcolor}
\usepackage{amsmath,amssymb}
\usepackage{braket}
\usepackage{graphicx}
\usepackage{stackengine}
\usepackage{bm}
\usepackage[utf8]{inputenc}
\usepackage{lmodern}
\usepackage{mathtools} 

\usepackage{tabularray}
\usepackage[mathlines]{lineno}
\usepackage{subcaption}
\usepackage{fancyhdr}
\usepackage{caption}
\usepackage{babel}
\usepackage{csquotes}
\usepackage{natbib}

\begin{document}
\preprint{APS/123-QED}

\title{Exploring the non-Markovian dynamics in depolarizing maps
}

 
\author{Ali  Abu-Nada$^{1}$}
\email{aabunada@hct.ac.ae}
\author{Subhashish Banerjee$^{2}$}
\email{subhashish@iitj.ac.in}
\author{Vivek Balasaheb Sabale$^{3}$}
\email{sabale.1@iitj.ac.in}
\affiliation{$^{1}$The Centre of Excellence for Applied Research and Training, Higher Colleges of Technology, Abu Dhabi, United Aarab Emirates\\}
\affiliation{$^{1}$ Khalifa Bin Zayed Air College, Al Ain, United Aarab Emirates\\}

\affiliation{$^{2}$ Department of Physics \\ Indian Institute of Technology Jodhpur, 342030, India\\} 

\affiliation{$^{3}$ Department of Chemistry \\ Indian Institute of Technology Jodhpur, 342030, India\\}
 
\date{\today}

\begin{abstract}
The non-Markovian depolarizing channel is explored from the perspective of understanding its non-Markovian behavior as well as the occurrence of singularities. The study brings together the various ways to identify and quantify non-Markovianity. This includes dynamical techniques such as quantum information backflow witness, Breuer-Laine-Piilo, Rivas-Huelga-Plenio and Hall-Cresser-Li-Andersson measures. In addition, geometrical visualization of non-Markovian effects is presented using the variation in the volume of accessible states during dynamical evolution. Further, a trajectory-based visualization of the dynamical map within the parameter space is presented. The trajectories traced during evolution demonstrate the loss of CP divisibility and the emergence of non-Markovianity under systematic variations of the system parameters.  The effects of increasing system dimensions and qubit numbers on singularity and non-Markovianity are presented, with an extension of characterization techniques to higher-dimensional systems. 

\end{abstract}

\maketitle


\section{Introduction }\label{sec:level1}

The dynamical evolution of the open quantum system \cite{breuer2002theory,banerjee2018open}, a composite system consisting of a smaller subsystem of interest and an interacting larger subsystem called environment, is of interest to the field of quantum information \cite{nielsen,thapliyal2017quantum} due to arising memory effects and non-Markovianity \cite{breuer,laine,rivas2,PhysRevA.99.042128,kumar2018non,utagi2020temporal,mixing,evolution}. The concept of the completely positive trace preserving (CPTP) map is used to study such evolution, which maps the systems initial state $\rho_{S}(0)$ to the evolved final state $\rho_{S}(t)$ \cite{nielsen,ruskai}
\begin{equation}\label{eq1}
     \rho_{S}(t) = \Phi(t,0) \rho_{S}(0).
 \end{equation}
Alternatively, CPTP maps are called quantum channels and can be described using operator sum representation  (Kraus formalism) \cite{choi1,choi2,kraus1,kraus2}. The mapped evolution by the CPTP map can be classified as Markovian or non-Markovian \cite{santis}. In Markovian evolution, changes in the system state depend solely on its current state, with no influence from past events. Conversely, in non-Markovian evolution, this independence is breached, indicating that the environment retains a memory of the system history, affecting its future evolution. \par
To identify non-Markovianity, traditionally, one resorts to the loss of CP divisibility \cite{rivas2} or nonmonotonic increase in the distinguishability of a pair of states during evolution \cite{laine}. Even CP-divisible processes can exhibit non-Markovian effects \cite{kumar2018non,utagi2020temporal}. The increase in distinguishability is indicative of information backflow from environment to system, indicating non-Markovianity. The Breuer-Laine-Piilo (BLP) measure \cite{breuer} uses distinguishability to quantify non-Markovianity.  However, the BLP measure is not the only method to detect non-Markovianity. There are other witnesses, such as the Rivas-Huelga-Plenio (RHP) measure \cite{rivas,rivas2} and Hall-Cresser-Li-Anderson (HCLA) measure \cite{hall}, as discussed below. The divisibility of a dynamical map can be used to characterize non-Markovianity. The dynamical map is said to be CP-indivisible, if the intermediate map (propagator), is not completely positive (NCP), possessing at least one negative eigenvalue for the corresponding Choi matrix. The intermediate NCP map is indicative of system-environment correlation. The RHP measure \cite{rivas,rivas2} uses the previously mentioned criteria to quantify non-Markovianity. Similarly, the HCLA measure \cite{hall} quantifies non-Markovian effects but relies on negative time-dependent decay rates, which are indicative of non-Markovianity. Both measures are equivalent to each other and provide similar insights into non-Markovian processes. \par

Here, we undertake a follow-up of \cite{shrikant} to further explore the non-Markovian depolarising map. The additional emphasis on the quantum nature of information backflow \cite{qi} and originating singularities is presented. The singularity structure originates in the intermediate map at the crossover of two eigenvalues and is one of the features of depolarizing maps. \par 

The study of non-Markovianity in quantum channels is necessary for understanding the quantum information processes. Non-Markovian dynamics can significantly affect the quantum systems ways of information preservation and manipulation. By studying non-Markovian effects, better protocols can be developed to mitigate errors \cite{wood,yanhu} in quantum systems, enhancing the reliability and capabilities of quantum computations and communications.

This paper adds to this growing body of research, addressing current gaps in understanding. Moreover, as quantum technology  \cite{ADEPOJU2017581,doi:10.1142/S0219749916500349} rapidly advances, the necessity of comprehending and understanding non-Markovian effects increases.

 Exploring non-Markovianity in open quantum systems is crucial in today's world, as modern technologies increasingly rely on non-Markovian processes. The importance of this study is driven by both foundational and practical considerations. Fundamentally, exploring the non-Markovian nature of system-environment interactions is essential for accurately describing a variety of real-world systems. Examples include quantum biological systems \cite{kenneth}, complex quantum networks \cite{Nokkala_2024}, solid-state devices such as superconductor quantum interference devices (SQUIDs) and Josephson junctions \cite{Khaneja19}, and ultracold gases \cite{bloch}, all of which exhibit non-Markovian behavior. Incorporating non-Markovian effects into the analysis of these physical systems allows for more precise and reliable predictions. 
 
In engineering and control systems \cite{reich}, incorporating non-Markovian dynamics can significantly enhance the performance and stability of control mechanisms. Similarly, in material science \cite{PhysRevResearch.5.L032037}, non-Markovian models play a crucial role in understanding and designing materials with tailored properties, as memory effects can influence their mechanical, thermal, and electrical characteristics, facilitating advancements in material design. By addressing both theoretical and practical aspects, the study of non-Markovian dynamics effectively bridges the gap between academic research and real-world applications, thereby advancing scientific understanding and fostering technological innovation.

The depolarizing channel is a fundamental noise model in quantum information theory, representing a type of noise that occurs in quantum systems. Analyzing its non-Markovianity can give insights into error correction \cite{Preskill1998,nielsen,shor}, quantum communication \cite{bassoli,meter}, and quantum computing \cite{Preskill1998,nielsen}.

The depolarizing channel can be produced by several types of environments, generally characterized by their randomizing effects on the quantum state. For example, in environments where the system is subjected to random fluctuations or noise, such as thermal noise, background radiation, or uncontrolled electromagnetic fields, the qubit can be depolarized \cite{bylicka,CHAPEAUBLONDEAU2022128300, PhysRevResearch.6.023263,martinez}. These fluctuations cause random rotations in the state of the qubit, leading to a mixture of all possible states, which is characteristic of the depolarizing channel. Furthermore, in optical quantum systems, depolarization can occur due to scattering processes. For example, when photons interact with particles in a medium, they can scatter in a way that randomizes their polarization, leading to a depolarized state \cite{dariano2003quantumtomography,PhysRevA.83.022303}. Thus, any environment that induces random, unbiased noise or interactions on a quantum system can give rise to a depolarizing channel. This randomness leads to the equal probability of all possible outcomes, which is the defining feature of depolarization.

Here, we try to establish a foundational understanding of non-Markovianity in the depolarizing channel. The simplicity of a one-qubit system allows for clearer interpretation and validation of theoretical models. Additionally, an extension of depolarizing noise to $N$-level qudit systems \cite{eltschka2021shape,hu2021novel,bertlmann2008bloch,PhysRevA.100.062311} and multiqubit systems is made, allowing for the study of non-Markovianity in higher-dimensional systems. \par
Our goal in this article is to answer the following questions: {\it{ (1) Under what conditions does the depolarizing map become non-Markovian? (2) Can we quantify the amount of this non-Markovianity? and (3) What will be the effect of the increased levels of system or number of qubits on non-Markovianity?}} In order to answer these questions, both dynamical (based on CP divisibility, distinguishability, negative time-dependent decay rates, and quantum information
backflow witness) and geometrical techniques are employed, and a comprehensive understanding of the non-Markovianity of depolarizing maps is attempted. The geometrical visualization involves descriptions of the non-Markovianity based on: an unexpected increase in the volume of accessible states of the system signifying the non-Markovian dynamics \cite{lorenzo}, and employing a trajectory-based visualization of traced trajectories within the parameter space to capture the dynamical evolution of the depolarizing channel, in turn demonstrating the loss of CP divisibility and the emergence of non-Markovianity under systematic variations in the system parameters \cite{filippov}.

This work is organized as follows. In Sec. \ref{sec:level2}, the general depolarizing map including the non-Markovianity parameters  is
derived. In Sec. \ref{sec:level3}, we describe the dynamical techniques in detecting the non-Markovianity of the depolarizing map by observing the negativity of  the eigenvalues of the Choi matrix of the intermediate map, witness for quantum information backflow, and the  canonical time-dependent decay rate. Furthermore, we utilize the  trace-distance-based distinguishability
method  to detect the non-Markovianity of the map. In Sec. \ref{level3}, we describe the geometrical tools in detecting the non-Markovianity through: (a) changing the volume of physical states of the system, and (b) employing
a trajectory-based visualization within the parameter space to trace trajectories and capture the
dynamical evolution of the depolarizing channel. In addition, we also characterize non-Markovianity in a qutrit system, thereby paving the way for understanding non-Markovian open quantum systems in higher dimensions, in Sec. \ref{sec:level4add}.  The non-Markovian behavior in multiqubit systems is presented in Sec. \ref{sec:level4new}.Finally, we conclude in Sec. \ref{sec:level5}.

\section{General depolarizing map}\label{sec:level2}

 The depolarizing map can be defined as a quantum operation that transforms the density matrix  of the initial  state of the system  
 into a convex combination of the original density matrix with probability of $1-k$ and the maximally mixed state  with probability $k$
 \cite{nielsen,shrikant}. Mathematically, it is expressed as $\Phi(\rho) = \sum_{i=\mathbb{I},X,Y,Z} E_{i} \rho E_{i}^{\dagger}$, where
\begin{align}\label{eq2}
E_{\mathbb{I}} = \sqrt{1-\frac{3}{4}k} \hspace{2mm} \mathbb{I}; \hspace{0.7cm} E_{X} = \sqrt{\frac{k}{4}} \hspace{2mm} \sigma_{X}, \nonumber \\
    E_{Y} = \sqrt{\frac{k}{4}} \hspace{2mm} \sigma_{Y}; \hspace{0.7cm}  E_{Z} = \sqrt{\frac{k}{4}} \hspace{2mm} \sigma_{Z},
\end{align}
where $\mathbb{I}$  is the identity operator; $\sigma_{X}$, $\sigma_{Y}$, $\sigma_{Z}$ are  the Pauli matrices; and $\sum_{i}E_{i}^{\dagger}E_{i}=\mathbb{I}$. Furthermore, $k$  increases monotonically from $0$ (noiseless case) to $1$  (maximal depolarizing) \cite{gupta}. The most general form of  the 
Kraus operators  of the depolarizing  map can be written as \cite{bylicka,shrikant,gupta}

\begin{align}\label{eq3}
     E_{\mathbb{I}} &= \sqrt{\left[1+\Upsilon_{1}(p)\right]\left(1-\frac{3}{4}p\right)}\hspace{2mm} \mathbb{I}; \nonumber \\ E_{X} &= \sqrt{\left[1+\Upsilon_{2}(p)\right]\frac{p}{4}} \hspace{2mm} \sigma_{X}; \nonumber \\
    E_{Y} &= \sqrt{\left[1+\Upsilon_{2}(p)\right]\frac{p}{4}} \hspace{2mm} \sigma_{Y}; \nonumber \\  E_{Z} &= \sqrt{\left[1+\Upsilon_{2}(p)\right]\frac{p}{4}} \hspace{2mm} \sigma_{Z},
\end{align}
where $\Upsilon_{1}(p)$ and $\Upsilon_{2}(p)$ are  real functions, and $p$ is a
time-like parameter that changes monotonically from $0$ to $1$. We regain  Eq. (\ref{eq2}) by adjusting $\Upsilon_{1}(p)= \Upsilon_{2}(p)=0$, and $p$  being replaced by $k$. 

To find out the form of  $\Upsilon_{1}(p)$ and  $ \Upsilon_{2}(p)$, we use the completeness condition

\begin{align} \label{eq4}   
    \sum_{i} E^{\dagger}_{i}E_{i}&= 
    \left(1+\Upsilon_{1}(p)\right)\left(1-\frac{3}{4}p \right) + \left(1+\Upsilon_{2}(p) \right) \left(\frac{3}{4}\right) p = 1\nonumber,\\ 
    & \implies \left(1-\frac{3}{4}p\right)\Upsilon_{1}(p) + \frac{3}{4}p \Upsilon_{2}(p)=0.    
\end{align}
This suggests that $\Upsilon_{1}(p) = -\frac{3}{4} \alpha p$ and $\Upsilon_{2}(p) = \alpha \left(1-\frac{3}{4}p \right)$, where  $\alpha$  is real number. The parameter $\alpha$ creates a small perturbation to the map and imposes non-Markovianity into the dynamics. Hence, it can be considered as a non-Markovian parameter.  

The general Kraus operators of the depolarizing map  then take the form

\begin{align}\label{eq5}
    E_{\mathbb{I}} &= \sqrt{\left[1-\frac{3}{4} \alpha p\right]\left(1-\frac{3}{4}p\right)}\hspace{2mm}\mathbb{I}; \nonumber \\ E_{X} &= \sqrt{\left[1+\alpha\left(1-\frac{3}{4}p \right)\right]\frac{p}{4}} \hspace{2mm}\sigma_{X}; \nonumber \\
    E_{Y} &=  \sqrt{\left[1+\alpha\left(1-\frac{3}{4}p \right)\right]\frac{p}{4}} \hspace{2mm}\sigma_{Y}; \nonumber \\  E_{Z} &=  \sqrt{\left[1+\alpha\left(1-\frac{3}{4}p \right)\right]\frac{p}{4}} \hspace{2mm} \sigma_{Z}.
\end{align}

To validate the map's complete positivity, we choose  $\alpha \in [0,1]$. For instance, if $\alpha =0$, then Eq. (\ref{eq5}) reduces to the Kraus operators of the standard depolarizing map, Eq. (\ref{eq2}), and  $p$ becomes $k$. In general, $k$ depends on $p$. Comparing
 Eq. (\ref{eq2}) with Eq. (\ref{eq5}), $k(p)$  can be expressed as 
\begin{equation}\label{eq6}
     k(p) = p +\alpha p-\frac{3}{4}\alpha p^{2}.
\end{equation}

\section{Non-Markovianity using dynamical tools}\label{sec:level3}

In this section, we characterize the non-Markovianity of the depolarizing map, given by Eq. (\ref{eq5}), by observing the negative eigenvalues of the Choi matrix of the intermediate map and the canonical time-dependent decay rate. In addition, the witness operator for non-Markovianity is discussed, and the quantification of resulting memory effects due to information backflow is also presented.  Furthermore, we utilize the  trace-distance-based distinguishability method to detect this non-Markovianity.

 If the dynamical map $\Phi(t,0)$ in Eq. (\ref{eq1}) is  CP divisible, then it can be represented as a series of propagators, described by CP maps $\Phi(t,s)$, $0 \leq s \leq t $,
\begin{equation}\label{eq7}
   \Phi(t,0) = \Phi(t,s) \circ \Phi(s,0).
\end{equation}

Furthermore, for invertible $\Phi (t,0)$ the propagator $\Phi(t,s)$ is well defined as \cite{rivas2,PhysRevLett.121.080407} 
\begin{equation}\label{eq8}
   \Phi(t,s) = \Phi(t,0) \left[ \Phi^{-1}(s,0)\right].
\end{equation}
Note that the condition in Eq. (\ref{eq7}) is the quantum counterpart to the classic Chapman-Kolgomorov equation \cite{rivas2}. However, if the intermediate map $\Phi(t,s)$ is not CP (NCP) \cite{jordan}, then it is indicative of the usual notion of non-Markovianity \cite{breuer,rivas2,shrikant}. 

The dynamical map $\Phi(t,0)$ in terms of ``divisibility" represents a memoryless evolution, as a composition of physical maps is indicative of quantum Markovianity. Based on the RHP criterion \cite{rivas2}, a dynamics is said to be non-Markovian if it is not CP divisible.  This does not require optimization but a normalization is needed to handle the singularity.  An alternative method of characterizing non-Markovian dynamics is the BLP method \cite{breuer,laine}. Here, non-Markovianity is defined 
in terms of the non monotonic behavior of the trace distance, $|| . ||_{1},$ and is based on the notion of distinguishability between any two initial states $\rho_1$ and $\rho_2$. 
A non monotonic behavior of distinguishability is indicative of a backflow of information from the environment to the system, a signature of non-Markovianity.  This requires optimization of the states and does not require normalization. The information backflow criteria does not distinguish between classical and quantum memory effects. Using \cite{qi}, quantumness in the observed information backflow can be characterized for the intermediate map.

\subsection{The eigenvalues of the   Choi matrix  of the intermediate dynamics \label{secIIA}}

Although the dynamical map $\Phi(t,0),t > 0$, can be obtained through tomography \cite{nielsen,rivas2}, the exponential scaling of this process makes it costly. However, detecting non-Markovianity does not necessarily require such procedures. By establishing accurate lower and upper bounds for properties such as  trace distance \cite{laine} or entanglement \cite{rivas2} using simpler measurements, we can detect nonmonotonic behavior without relying on expensive tomography. Also, advancements in the field of open quantum systems report possible characterization of dynamical maps even for higher dimensions \cite{bouchard2019quantum,PhysRevA.107.042403,PRXQuantum.3.020344}. For the present case, the general depolarizing map is theoretically known; see Eq. (\ref{eq5}). 
To realize the non-Markovianity of the  general depolarizing  map given by Eq.  (\ref{eq5}), we consider the intermediate map $\Phi(p,q)$, $0 \leq q\leq p$. If the intermediate map $\Phi(p,q)$ is not CP, the dynamics will be non-Markovian \cite{rivas2}.
 
In the following,  we describe the main steps toward calculating the intermediate map in Eq. (\ref{eq8}), $\Phi(p,q)$. The  general depolarizing map is written as

\begin{align}\label{eq9}
\Phi(p,0) (\rho(0)) &= E_{\mathbb{I}} \rho(0) E^{\dagger}_{\mathbb{I}} + \sum_{i=X,Y,Z} E_{i} \rho(0) E_{i}^{\dagger}  \nonumber\\
 &= 
\left[1-\frac{3}{4} \alpha p\right]\left(1-\frac{3}{4}p\right) \hspace{1mm} \rho(0)+ \nonumber \\
& \sum_{i = X,Y,Z} \left[1+\alpha\left(1-\frac{3}{4}p \right)\right]\frac{p}{4}
\hspace{1mm}\sigma_{i}\rho(0) \sigma_{i}.
\end{align}
Furthermore, we vectorize  $\Phi(p,0)$ by writing  the resulting matrix as a (column) vector by gathering the columns on top of one another. This process is  called ``vectorization"  \cite{rivas2,horn,magnus}.

The diagonalized form of the matrix elements of the {\it{unital}} intermediate map obtained using $\Phi(p,0)$, following  Eq. (\ref{eq8}) is
\begin{equation}\label{eq10}
\Phi(p,q) = \Phi(p,0)  \Phi^{-1}(q,0) = 
\begin{pmatrix}
1&0&0&0\\
0&\lambda_{1}(p)&0&0\\
0&0&\lambda_{2}(p)&0\\
0&0&0&\lambda_{3}(p)
\end{pmatrix},
\end{equation}
where,
\begin{align}\label{eq11}
   \lambda_{1}(p)= \lambda_{2}(p) = \lambda_{3}(p) = \frac{p\left(4+4 \alpha -3 \alpha p \right)-4} {4q+4 \alpha q - 3 \alpha q^{2}-4}.
\end{align}
Here, $0 \leq q  \leq p \leq 1$,   where $p$ and $q$ are regarded as 'timelike' variables, in place of using time $s$ and time $t$. 
It is worth mentioning here that when $q=\alpha=0$ in Eq. (\ref{eq11}), then $\Phi(p,q) =\Phi(p,0)$, and $\lambda_{1}(p) = \lambda_{2}(p) = \lambda_{3}= (1-p)$. This illustrates  the standard  depolarizing  map, and  the   map   shrinks the Bloch sphere uniformly  along $x$, $y$, and $z$ to a radius of 
$(1-p)$ \cite{lidar,Preskill1998}.

To find out whether or not $\Phi(p,q)$ is NCP, we use a Choi matrix constructed by first constructing  the matrix $U_{2\leftrightarrows 3}[\Phi(p,q) \otimes  \mathbb{I}_{4}]U_{2\leftrightarrows 3}$ where $U_{2\leftrightarrows 3}$  is the commutation
(or ``swap") matrix \cite{horn1994topics,magnus} between the ``second" and the ``third" subspace \cite{rivas2}; second,
apply $U_{2\leftrightarrows 3}[\Phi(t,s) \otimes  \mathbb{I}_{4}]U_{2\leftrightarrows 3}$ on  $\rm{vec}(\ket{\Psi}\bra{\Psi})$ (vectorization of $\ket{\Psi}\bra{\Psi}$), where $\ket{\Psi}= \frac{1}{\sqrt{2}} \left( \ket{00} +\ket{11} \right)$.  $U_{2\leftrightarrows 3}$ is written as follows
\begin{equation}\label{eq12}
U_{2\leftrightarrows 3} = \mathbb{I}_{2} \otimes
\begin{pmatrix}
1&0&0&0\\
0&0&1&0\\
0&1&0&0\\
0&0&0&1
\end{pmatrix}\otimes \mathbb{I}_{2}.    
\end{equation}

We assume that $\phi(p,q) \otimes  \mathbb{I}_{4}$ acts on the tensor product of four
spaces with the same dimension, $\mathcal{H}_{1}\otimes\mathcal{H}_{2} \otimes \mathcal{H}_{3} \otimes\mathcal{H}_{4}$. Then operator $U_{2\leftrightarrows 3}$ is the permutation matrix causing the interchange of the second and third subspace. Third, we write the result as a matrix, $i.e.$ “devectorize” to construct the Choi matrix  of the intermediate map as follows

\begin{equation}\label{eq13}
    \chi(\alpha, q,p)  = \left[ \Phi(p,q) \otimes \mathbb{I}  \right]\ket{\Psi} \bra{\Psi}.     
\end{equation}
The generalization of this process for the $N$-level system is presented in the Appendix \ref{sec:appendix}.
By Choi-Jamiolkowski isomorphism, matrix $\chi(\alpha, q,p)$ is
positive if and only if $\Phi(p, q)$ is CP \cite{shrikant,omkar}. A nonpositive semidefinite Choi matrix of propagator indicates non-Markovianity. The  Choi
matrix for the intermediate map, $\chi(\alpha, q,p)$, is found to be
\begin{widetext}
\begin{equation}\label{eq14}
\chi(\alpha, q,p)=
  \begin{pmatrix}
\frac{p+q+\alpha (p+q)-\frac{3}{4} \alpha (q^{2}+p^2)-2}{q+\alpha q -\frac{3}{4} \alpha q^{2}-1}&0&0&\frac{(1+\alpha)p- \frac{3}{4} \alpha p^{2}-1}{2(1+\alpha)q -\frac{3}{2}\alpha q^{2} -4}\\

0&\frac{(q-p)\left(\alpha -\frac{3}{4} (p+q) + 1\right)}{q\left(1+\alpha\right) -\frac{3}{4} \alpha q^{2}-1}&0&0\\

0&0&\frac{(q-p)\left(\alpha -\frac{3}{4} (p+q) +1 \right)}{q\left(1+\alpha\right) -\frac{3}{4} \alpha q^{2}-1}&0\\

\frac{(1+\alpha)p- \frac{3}{4} \alpha p^{2}-1}{2(1+\alpha)q -\frac{3}{2}\alpha q^{2} -4}&0&0&\frac{p+q+\alpha (p+q)-\frac{3}{4} \alpha (q^{2}+p^2)-2}{q+\alpha q -\frac{3}{4} \alpha q^{2}-1}
\end{pmatrix}.
\end{equation}
\end{widetext}

\begin{figure}[htpb]
   \includegraphics[width=9cm]{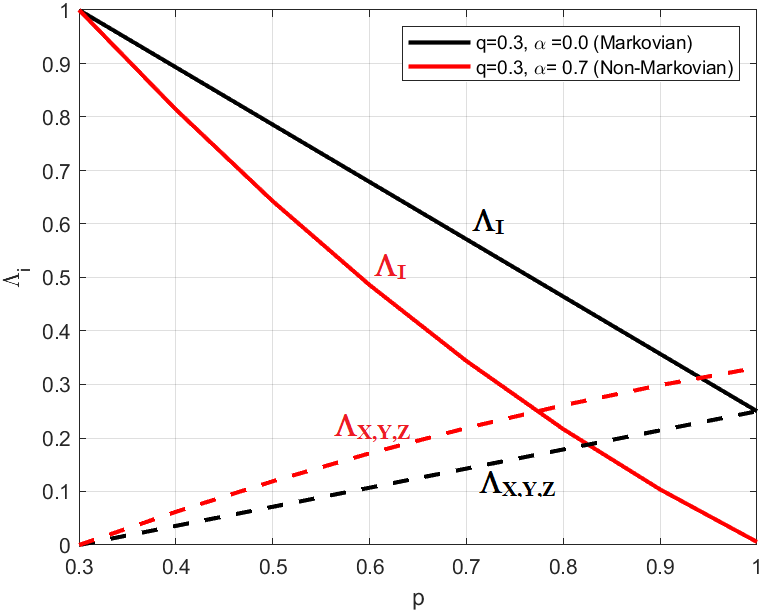}
    \caption{(Color online). The  eigenvalues of the Choi
matrix (\ref{eq14}). Solid-black line ($\Lambda_{\mathbb{I}}$) and dashed black line ($\Lambda_{X,Y,Z}$), both for $\alpha=0.0$. Solid red line ($\Lambda_{\mathbb{I}}$) and red dashed line ($\Lambda_{\mathbb{I}}$) both for $\alpha=0.7$. The intermediate $p$ range  lies between $ p=q=0.3$ and $p=p_{max}=1.0$. Here,$q<\alpha_{-}$, and crossover at $\alpha_{-}\approx 0.78$ .}
    \label{fig:fig1}
\end{figure}

The  eigenvalues of $\chi(\alpha, q,p)$ are
\begin{align}\label{eq15}
    \Lambda_{\mathbb{I}} &= \frac{1}{4} + \frac{3}{4} \left[\frac{p\left(4+4 \alpha -3 \alpha p \right) -4} {4 q +4 \alpha q -3 \alpha q^{2}-4} \right]; \nonumber \\
     \Lambda_{i} &= \frac{1}{4} - \frac{1}{4} \left[\frac{p\left(4+4 \alpha -3 \alpha p \right) -4} {4 q +4 \alpha q -3 \alpha q^{2}-4} \right],
\end{align}
where $i = \{X,Y,Z \}$. In the case of no perturbations, $i.e.,$ $\alpha=0$, and $q=0$, the map is CP-divisible as it reduces to a simple case of a Markovian depolarizing channel. In this case, the eigenvalues become $\Lambda_{\mathbb{I}} = \left(1-\frac{3}{4}p\right)$ and $\Lambda_{i} = (\frac{p}{4})$. The same is true for $\alpha\neq0$ and $q=0$, just presenting the perturbed depolarizing channel. Furthermore, when $p=p_{max}=1.0$, the  depolarizing  is maximum,
indicated by the intersecting points of the black lines in Fig. \ref{fig:fig1}.

\begin{figure}[htpb]
    \centering
    \includegraphics[width=9cm]{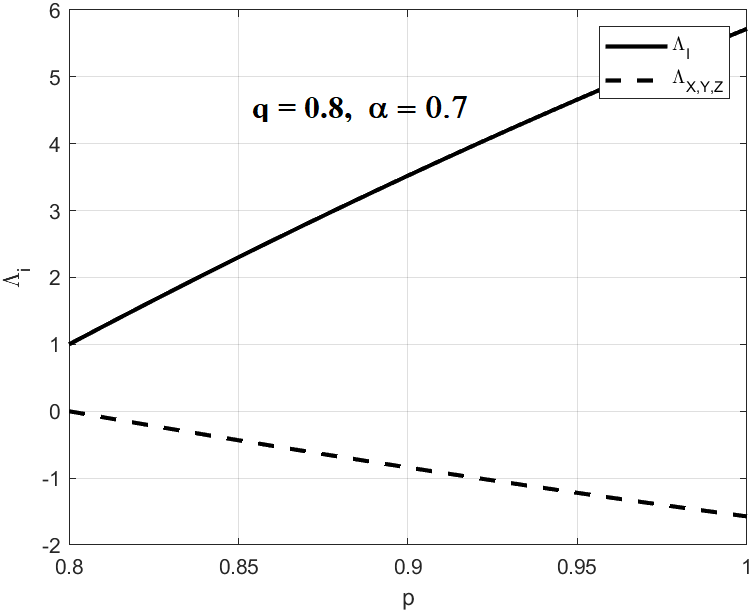}    
\caption{ The  eigenvalues of the Choi matrix (\ref{eq15}). Sold line ($\Lambda_{\mathbb{I}}$) and,   dashed line ($\Lambda_{X,Y,Z}$) , for $\alpha=0.7$ (non-Markovian).   The intermediate $p$ range  lies between $ q=0.8$ and $p_{max}=1.0$. The whole range of $p$  corresponds to an NCP map, demonstrating the non-Markovianity of the channel. Here $q>\alpha_{-}$ as for $\alpha=0.7$ corresponding with $\alpha_{-}\approx0.78$.}
\label{fig:fig2}
\end{figure}

On the other hand, introducing a perturbation during the dynamics, initiates some form of memory into the system. These memory effects are controlled by the factor $\alpha$. The eigenvalues of the Choi matrix show crossover for $\alpha\neq0$ and  $q\neq0$. The timelike $p$ varies from $q$ to $p_{max}=1$. The $p$ satisfying $4 p +4 \alpha p -3 \alpha p^{2}-4=0$ gives $\alpha_{\pm}=\frac{2(1+\alpha \pm \sqrt{1-\alpha+\alpha^{2}})}{3\alpha}$, corresponding to points of crossover of eigenvalues. The $\alpha_{+}$ is outside the domain of $p$ and is not considered for study. For $q < \alpha_{-}$, eigenvalues crossover at $\alpha_{-}$. As depicted in Fig.  \ref{fig:fig1}, the point $p=\alpha_{-}$ represents maximal depolarizing. 
For $q > \alpha_{-}$, as illustrated in Fig. \ref{fig:fig2}, $\Lambda_{i} $  become negative in the entire range of $p$ for the intermediate map indicating  non-Markovianity. The point $q=\alpha_{-}$ represents a singularity as the propagator or intermediate map is undefined at this point; the $\Lambda_{\mathbb{I}}$ and  $\Lambda_{X,Y,Z}$ diverge for any $p \in (q, 1]$. The negative eigenvalue  $\Lambda_{X,Y,Z}$ suggests that the trace norm of the Choi matrix, $||\chi(\alpha, q, p)||_{1}$ is greater than 1 in this interval. 
 Since $\Phi(p,q)$  is trace preserving, 
 $||\chi(\alpha,q,p)||_{1}$ provides a  witness of the NCP character of $\Phi(p,q)$ as \cite{rivas2}

\begin{align}\label{eq16}
||\chi(\alpha, q, p)||_{1}&=||\left[ \Phi(p,q) \otimes \mathbb{I}  \right]\ket{\Psi} \bra{\Psi}||_{1}\nonumber \\
&\begin{cases}
      =1, \text{if }  \hspace{2mm} \Phi(p,q)\hspace{2mm}  is\hspace{2mm}  CP\hspace{2mm} \\
      >1,  \text{if }  \hspace{2mm} \Phi(p,q)\hspace{2mm} is \hspace{2mm} NCP.     
    \end{cases}
\end{align}

Integrating $\chi(\alpha,q, p)$ over the evolution time provides a measure of non-Markovianity, $i.e.,$ the RHP measure \cite{rivas2}. This quantifies non-Markovian effects in the CP-indivisible regime.
Moreover, the NCP intermediate map relates to negative decay rates, suggesting a conceptually equivalent, but
quantitatively different and possibly computationally easier method to indicate non-Markovianity, based on the integral of the normalized time dependent decay rate function in the canonical form of the master equation for the negative decay rate(s),  $i.e.,$ the HCLA measure \cite{hall}, used below. 

\subsection{ The negativity of the decay rate\label{sec:HCLA}}

As stated in the preceding subsection,  an NCP
intermediate map is related to the observation of the negativity of the time-dependent  decay rate $\gamma(p)$ of the master equation in the canonical form. 

In some cases of interest, the open system dynamics may be written in terms of a time local master equation involving time-dependent functions as prefactors with otherwise Gorini-Kossakowski-Sudarshan-Lindblad (GKSL) form \cite{gorini,lindblad}. Then, for some  times during the system evolution, negative decay rates may show up, indicative of the  non-Markovianiaty of the dynamics. The generalized Markovian dynamics appears when the master equation takes the quasi-GKSL-form \cite{nina}
\begin{align} \label{eq17}
\frac{d \rho(p)}{dp}&=-i\left[H,\rho(p) \right]+ \sum_{i}    \gamma_{i}(p) \left( L_{i} (p) \rho(p) L_{i}^{\dagger}(p) \right. \nonumber \\ 
&\left. -\frac{1}{2}  \left\{ L_{i}^{\dagger}(p) L_{i} (p), \rho(p) \right\}\right ),
\end{align}    
 written in a canonical form \cite{hall}, where the $L_{i}(p)$ are traceless orthonormal operators. The  decay rates $\gamma_{i}(p) \geq 0$ defines a  CP divisible dynamical map. It is natural to regard dynamical maps $\Phi(t,0)$ with master equations of the type given by  Eq. (\ref{eq17}) for which $\gamma_{i}(p) < 0$ for some $i$ and some time as candidates for non-Markovian quantum dynamics \cite{nina}.  In these cases, the dynamical map is no longer CP divisible. 
\begin{figure}[htpb]
    \centering
    \includegraphics[width=9cm]{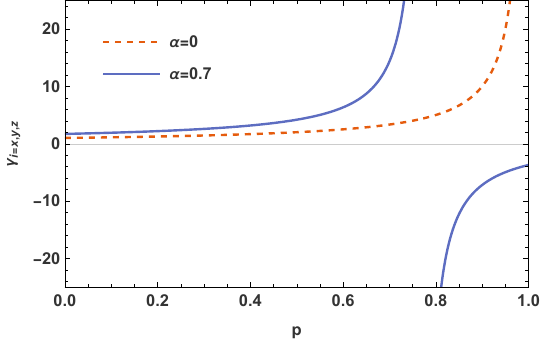}
\caption{ (Color online) A plot of decoherence rates, $\gamma_{X,Y,Z}(p)$, as a function of $p$ for $\alpha =0.0$ (red dashed curve) and $\alpha=0.7$ (blue solid curve).}
 \label{fig:fig3}
\end{figure}

The Kraus representation given by Eq. (\ref{eq5}) is a solution to the master equation describing the depolarizing map in the canonical form:
\begin{equation}\label{eq18}
    \frac{d \rho}{dp} = \sum_{i=X,Y,Z} \gamma_{i}(p) \left[\sigma_{i} \rho(p) \sigma_{i} -\rho(p)\right],
\end{equation}
where $\gamma_{X}=\gamma_{Y}=\gamma_{Z} $ and  $\int_{0}^{p} \gamma_{i}(p) dp > 0$ for CP dynamics. We anticipate that the decay rates $\gamma_{i}(p)$ become negative, from $ p=0.8$ to $p= 1 $, and thus indicating the non-Markovianity in this range. To calculate $\gamma_{i}(p)$, we let $G=1-k(p)$, see [Eq. (\ref{eq6})], and use the formula \cite{david}

\begin{equation}\label{eq19}
    \gamma (p) = -\frac{1}{G}\frac{dG}{dp}= \frac{4+(4-6 p)\alpha}{4 + 3 \alpha p^{2}-4p(1+\alpha)}.
\end{equation} 
Figure \ref{fig:fig3} illustrates a plot of decay rates $\gamma_{X,Y,Z}(p)$ as a function of $p$ for $\alpha =0.0$ (red dashed curve) and $\alpha=0.7$ (blue solid curve). We observe that when the non-Markovian (perturbation) parameter $\alpha$ is equal to zero, the decay rates become positive for the entire range of $p$ and then  the dynamics  is CP divisible, as expected. Furthermore, when  $\alpha= 0.7$,  the decay rates turn into negative at the singularity point, $i.e.,~ p \approx 0.78$,  indicating  non-Markovian dynamics.

\begin{figure}[htpb]
    \centering
    \includegraphics[width=9cm]{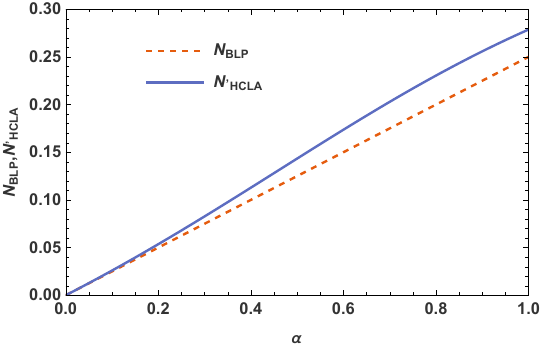 }
\caption{ A plot of the $BLP$ measure $\mathcal{N}_{BLP}$ (red dashed line) and $HCLA$ measure $\mathcal{N}^{'}_{HCLA}$ (blue solid curve). 
 }
\label{fig:fig4}
\end{figure}

Using the canonical decay rates $\gamma_{i}(p)$  and  the master equation, given by Eq. (\ref{eq18}), a function $f_{i} (p)$ is defined \cite{hall},
\begin{equation}\label{eq20}
    f_{i} (p) = \max[- \gamma_{i}(p),0]\geq 0.
\end{equation}
If  $f_{i}(p)=0$,  at any time, then the evolution is CP divisible. The functions $f_{i} (p)$ can be used to  quantify the  non-Markovianity as
\begin{equation}\label{eq21}
 \mathcal{N}_{HCLA} = \int_{p=q}^{p=p_{max}} -\gamma(p) dp, 
\end{equation}
where $\gamma (p)$ is as in Eq. (\ref{eq19}). It diverges because  of the singularity  of $\gamma(p)$ at $p\approx 0.8$ (see Fig. \ref{fig:fig3}). One solution  to this problem, following an idea proposed in \cite{rivas2,shrikant}, is to replace $\gamma(p)$ by its normalized form

\begin{align}\label{eq22}
\tilde{\gamma}(p) &=  \frac{-\gamma(p)}{1-\gamma(p)}\nonumber \\
&= \frac{4+4\alpha -6 \alpha p }{4p+4\alpha -2 \alpha p   - 3 \alpha p^{2}  }.
\end{align}
Hence, 
\begin{align}
    \label{eq23}
    \mathcal{N}^{'}_{HCLA} &= \int_{\alpha_{-}}^{1}\tilde{\gamma} 
    (p) dp \nonumber \\
    &=  \Biggl\{ \ln \left(2p(-2+\alpha)-4\alpha+3p^{2}\alpha\right) \nonumber \\
    &  +\frac{6 \alpha ( \tanh^{-1}\left(\frac{-2+\alpha +3p\alpha}{\sqrt{4-4\alpha+13\alpha^{2}}}\right))}{\sqrt{4-4\alpha+13\alpha^{2}}} \Biggr\}^{p=1}_{p=\alpha_{-}}.
\end{align}
Figure \ref{fig:fig4} represents  the values of $\mathcal{N}^{'}_{HCLA}$ (blue solid curve) as a function of non-Markovian (perturbation) parameter $\alpha$. The monotonic increase of this measure with $\alpha$ indicates non-Markovianity. These results are connected to the RHP measure, $\mathcal{N}_{RHP}$, 
since both coincide for the two level system \cite{hall}. 

\subsection{The distinguishability of quantum states }

The  BLP method observes the distinguishability between  two initial states,  $\rho^{S}_{1}$ and $\rho^{S}_{2}$ by calculating  the $trace$ $distance$, $D$, between them. $D$  is a metric quantity on the space 
of physical states, satisfying $0 \leq D \leq 1$, where $D=1$ if and only if $\rho^{S}_{1}$ and $\rho^{S}_{2}$ have orthogonal support, and characterizes
a realistic upper bound for the distinguishability between the probability distributions resulting  from measurements executed on the quantum states. Therefore, the
trace distance can be given an interpretation as the distinguishability between two quantum states \cite{breuer,nielsen}. 
If $\Phi(p,0)$ is ``CP divisible,"  then it reflects as a contraction for the trace distance, $i. e.,$ $D$ is nonincreasing under CP divisible maps,
\begin{equation}\label{eq24}
  D \left[\Phi(p,0)(\rho^{1}_{S}), \Phi(p,0)(\rho^{2}_{S} )\right] \leq D\left[ \rho^{1}_{S}, \rho^{2}_{S}\right],
\end{equation}
where, in general,
\begin{equation}\label{eq25}
    D\left[ \rho^{2}_{S}, \rho^{1}_{S}
  \right] = \frac{1}{2} || \rho^{2}_{S}-\rho^{1}_{S} ||_{1}= \frac{1}{2}\text{tr}\sqrt{(\rho^{2}_{S}-\rho^{1}_{S})^{2}}.
\end{equation}
The violation of Eq. (\ref{eq24}) would be an indicator of non-Markovian behavior \cite{breuer}.  This is a sufficient but not necessary condition. The non-Markovian dynamics satisfying this violation is of P-indivisible nature \cite{rivas2}. The violation given by Eq. (\ref{eq24}) witnesses non-Markovian effects which are popularly interpreted as information backflow. The information backflow can be quantum as well as classical in nature. We justify, below, the quantumness of information backflow observed in the depolarizing channel, using the recently proposed memory witness in \cite{qi}.

 As an example, we calculate the trace distance between two different initial states, with orthogonal support, which evolve under depolarizing  dynamics, given by Eq. (\ref{eq9}). Let 
$\rho^{1}_{S}(0) = \frac{1}{2}
\begin{pmatrix}
1& -1\\
-1&1
\end{pmatrix},$
and $\rho^{2}_{S}(0) = \frac{1}{2}
\begin{pmatrix}
1& 1\\
1&1
\end{pmatrix}$. By applying the depolarizing channel, given by Eq. (\ref{eq9}), on these initial states, the final states turn out to be
\begin{align} \label{eq26}
    \rho^{1}_{S}(t)  &=\Phi(p,0) \left[\rho_{S}^{1}(0) \right]  \nonumber \\
    &=\frac{1}{2} \begin{pmatrix}
\frac{1}{2}& \frac{p}{2} + \frac{\alpha p}{2} - \frac{3}{8}\alpha p^{2} - \frac{1}{2}\\
\frac{p}{2} + \frac{\alpha p}{2} - \frac{3}{8}\alpha p^{2} - \frac{1}{2}&\frac{1}{2}
\end{pmatrix}, \nonumber \\
 \rho^{2}_{S}(t) &= \Phi(p,0) \left(\rho_{S}^{2}(0) \right) ]   \nonumber \\
    &=\frac{1}{2} 
    \begin{pmatrix}
\frac{1}{2}& -\frac{p}{2} - \frac{\alpha p}{2} + \frac{3}{8}\alpha p^{2} + \frac{1}{2}\\
-\frac{p}{2} - \frac{\alpha p}{2} + \frac{3}{8}\alpha p^{2} + \frac{1}{2}&\frac{1}{2}
\end{pmatrix}.
\end{align}
By substituting  Eq. (\ref{eq26})  into Eq. (\ref{eq25}), we obtain
\begin{align}\label{eq27}
 D\left[ \rho^{2}_{S}, \rho^{1}_{S}  \right] &= \frac{1}{2}\text{tr}\sqrt{(\rho^{2}_{S}-\rho^{1}_{S})^{2}}\nonumber \\
 &=\frac{1}{4} | 4+ 3\alpha p^{2} -4p(\alpha +1) |.
\end{align}
\begin{figure}[htpb]
    \centering
    \includegraphics[width=9cm]{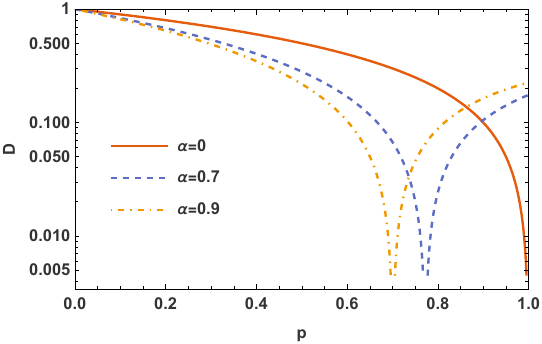}
\caption{  Logarithmic plot of trace distance $D$  for $\alpha=0.0$ (solid red curve), $\alpha=0.7$ (blue dash curve), and  $\alpha =0.9$ (orange dashed-dotted curve),  as a function of $p$. Note that, $\alpha=0.0$ describes the Markovian dynamics since $D$ decreases monotonically for all values of $p$. }
\label{fig:fig5}
\end{figure} 
The logarithmic plot of trace distance $D$  as a function of $p$ with $\alpha = 0$ (solid red curve), $\alpha = 0.7$ (blue dashed curve), and $\alpha =0.9$ (orange dash dotted curve), under the considered non-Markovian depolarizing  noise is depicted in  Fig. \ref{fig:fig5}. 
Note that larger $\alpha$ shows a larger enhancement region, suggesting larger
non-Markovianity in the sense of BLP \cite{breuer}.
The standard form of the BLP measure of non-Markovianity, $\mathcal{N}_{BLP}$, is given 
by

\begin{equation}\label{eq28}
\mathcal{N}_{BLP}  = \max_{\rho_{1},\rho_{2}} \int \frac{dD}{dp}  dp.
\end{equation}

If $\frac{dD}{dp} > 0 $ for some $p$, then the dynamics is non-Markovian \cite{rivas2} and the quantification of this non-Markovianity can be done by calculating $\mathcal{N}_{BLP}$. Equation (\ref{eq28}) involves maximizing distinguishability across all possible initial states.  Recall,  the initial states $\rho^{1}_{S}$ and $\rho^{2}_{S}$ in our example possess orthogonal support. Consequently, their distinguishability is at its maximum, which is 1. Hence, the BLP measure is given as
\begin{align}\label{eq29}
    N_{BLP}=  \int _{\alpha_{-}}^{1.0} \frac{dD}{dp}dp =\frac{\alpha}{4} . 
\end{align}
$\mathcal{N}_{BLP}$ is illustrated in Fig. \ref{fig:fig4}  (red dash line) and shows an agreement
with the quantification of non-Markovianity according to the normalized $\mathcal{N}^{'}_{HCLA}$.


\subsection{Using quantum memory witness}

\begin{figure}[htpb]
    \includegraphics[width=9cm]{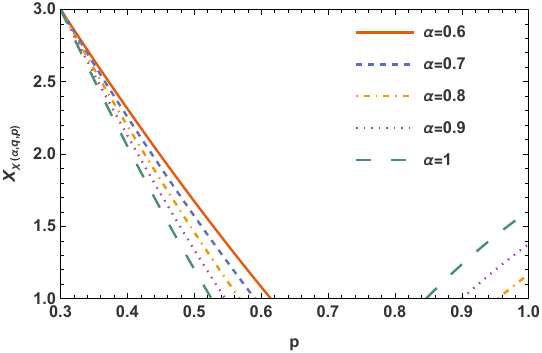}
\caption{ The variation of $X_{\chi(\alpha, q, p))}$, given by Eq. (\ref{eq32}), for $q=0.3, \alpha=0.$ (red solid line), $q=0.3, \alpha=0.7$ (blue dashed line), $q=0.3, \alpha=0.8$ (orange dash dotted line), $q=0.3, \alpha=0.9$ (purple dashed line), and $q=0.3, \alpha=1$ (green large-dashed line).}
 \label{fig:fig6}
\end{figure}

The intermediate dynamical map $\Phi(p,q)$ and the corresponding Choi matrix $\chi(\alpha, q, p)$ are useful for the observation and certification of non-Markovian memory effects. The breaking of CP divisibility can be captured using the Choi matrix, which is Hermitian, allowing one to write a spectral decomposition as $\chi(\alpha, q, p)=\sum_{i}\lambda_{i}P_{i}$. The $P_{i}$ are orthogonal projections derived using corresponding eigenvectors, and the $\Lambda_{i}$ are eigenvalues associated with $\chi(\alpha, q, p)$. The occurrence of negative $\Lambda_{i}$ points to non-Markovianity. Following this, the Bell states $(\ket{\phi^{+}},\ket{\psi^{\pm}})$ are observed to serve as witness operators for the depolarizing channel \cite{qi2} giving 
\begin{align}\label{eq30}
    \text{tr}(\ket{\phi^{+}}\bra{\phi^{+}}\chi(\alpha,q,p))= \Lambda_{\mathbb{I}}, \nonumber \\
    \text{tr}(\ket{\psi^{\pm}}\bra{\psi^{\pm}}\chi(\alpha,q,p))= \Lambda_{i=x,y,z}.
\end{align}
One of the eigenvalues ($\Lambda_{\mathbb{I}},\Lambda_{i=x,y,z}$) becomes negative for $q=0.3$ and $\alpha>0.7$, indicating that the intermediate evolution is not CP divisible in the interval [$q,p$] and, hence non-Markovian. The map admits information backflow in this interval and the usual tools lack the ability to distinguish this information as classical or quantum in nature. We use Ref. \cite{qi} to characterize the quantum nature of information backflow in the intermediate depolarizing map. The existence of the Hermitian witness operator, in the form of the Bell states, then allows the establishment of the variable $X_{\chi(\alpha, q, p)}$ corresponding to $\chi(\alpha, q, p)$. This is used to ascertain the quantum nature of the information backflow, characterized by nonmonoticity of $X_{\chi(\alpha, q, p)}$, by satisfying the following condition \cite{qi}
\begin{equation}\label{eq31}
   X_{\chi(\alpha, q, p)}= |\vec{s}| + ||T||_{1} >1,
\end{equation}
where $\vec{s}$ consists of $s_{i}= \text{tr}(\chi(\alpha, q, p)(\mathbb{I}\otimes \sigma_{i}))$, and $T$ is a 3$\times$3 matrix consisting of elements $T_{ij}= \text{tr}(\chi(\alpha, q, p)(\sigma_{i}\otimes \sigma_{j}))$. The expression for $X_{\chi(\alpha, q, p)}$ is  
\begin{equation}\label{eq32}
    X_{\chi(\alpha, q, p)}= 3  \frac{ \lvert p\left(4+4 \alpha -3 \alpha p \right) -4 \rvert} {\lvert 4 q +4 \alpha q -3 \alpha q^{2}-4 \rvert}.
\end{equation}
It quantifies the memory effects, and any nonmonotonic increase in $X_{\chi(\alpha, q, p)}$ can be ascribed to quantum information backflow in interval [$q,p$].
 Fig. \ref{fig:fig6} depicts the variation of quantum memory effects using $X_{\chi(\alpha, q, p)}$. For $q=0.3$ and $\alpha=0.8$, Fig. \ref{fig:fig6} depicts a non-monotonic increase, implying information backflow which is quantum in nature and not classical \cite{qi}. Further nonmonotonic variation of $X_{\chi(\alpha, q, p)}$ increases as $\alpha$ increases, indicating enhanced quantum information backflow and, hence non-Markovianity. The increase in value of $X_{\chi(\alpha, q, p)}$ is due to the system recovering information, particularly quantum information lost to the environment.

\begin{figure*}[htpb]
	\centering
	\begin{minipage}{1\columnwidth}
		\centering
		\includegraphics[width=8.7 cm]{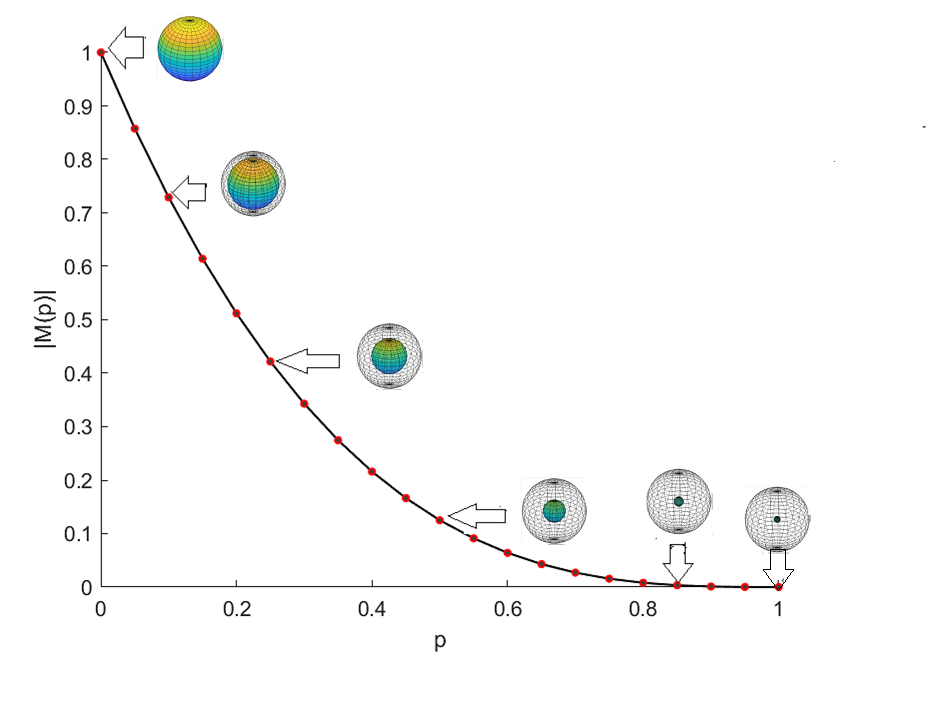}
	\end{minipage}%
	\begin{minipage}{1\columnwidth}
	\centering		\includegraphics[width=9cm]{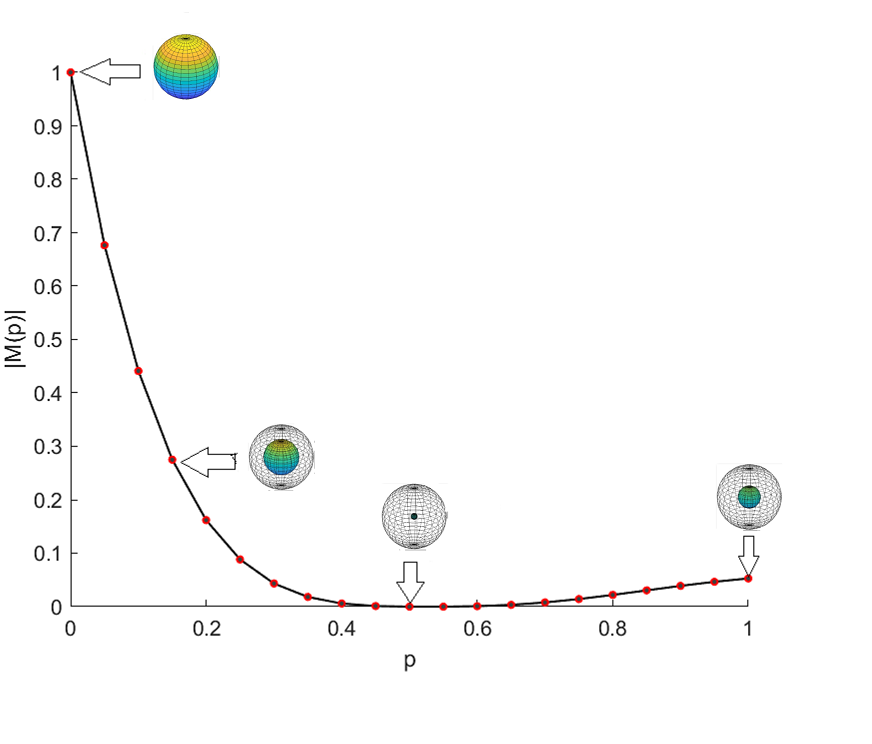}
	\end{minipage} 
 \caption{Time evolution of the  $|M(p)|$ for a generic Markovian (left graph) and non-
Markovian  (right graph) dynamics. In the Markovian case, the Bloch sphere shrinks along the $x$, $y$, and $z$ axes and so the  volume decreases monotonically for all values of $p$. For the non-Markovian case, the volume  suddenly grows at $p \approx 0.7$ indicating the non-Markovianity beyond this point. }
 \label{fig:fig7}
\end{figure*}


\section{Non-Markovianity via Geometrical visualization techniques}\label{level3}
In this section, we demonstrate the geometrical observation of the non-Markovian behavior of the depolarizing map through witnessing and visualizing the varying of the rate of the volume of physical states of the system affected by the depolarizing dynamics. Furthermore, we visualize  the divisibility of the depolarizing map and the conditions for the non-Markovian dynamics of the depolarizing channel.

\subsection{Non-Markovianity through volume change
of physical states of the system}\label{sec:level6}

The rate of change in the volume of accessible states to the  system   is a tool in Ref. \cite{lorenzo}, enabling a geometrical visualization of the dynamical effects of non-Markovianity. Unlike other methods (see, for example,  \cite{breuer}), this method  need not  be optimized over the initial states. If the  quantum evolution is Markovian, then this implies that the domain’s volume of the dynamical map reduces  monotonically, as indicated in the left side of Fig. \ref{fig:fig7}. In contrast, for a non-Markovian dynamical map, there could be an increase in the  domain’s volume for some time interval during the dynamics. 

The density matrix of a qubit can be written using
operators $G=\frac{1}{\sqrt{2}}\{ \mathbb{I},\sigma_{X}, \sigma_{Y}, \sigma_{Z}\}$ and generalized Bloch vector ($\Vec{r}$) \cite{nielsen,lidar1} as 
\begin{equation}\label{eq33}
    \rho= \sum_{i=0}^{3}r_{i}G_{i}.
\end{equation}
Here, $r_{i}=\text{tr}(G_{i}\rho)$ such that $\vec{r}=(\frac{1}{\sqrt{2}},r_{i=1,2,3})$.
If a map $\Phi$ acts on a single qubit then $\Phi: \rho \rightarrow \rho^{\prime}$, where $\rho^{\prime}$ must be expressible in terms of a new Bloch vector $\vec{r}^{\prime}$ associated to $\rho^{\prime}$. The action of a {\it unital} quantum map $\Phi$ on a qubit state can be represented by mapping the Bloch vector according to $\vec{r} \rightarrow \vec{r}^{\prime} = M \vec{r}$. 
The equation, $M_{ij}=\text{tr}[G_{i}\Phi[G_{j}]]$ gives the respective matrix elements. 
This is an affine transformation for the Bloch vector, for the respective $M$, as

\begin{equation} \label{eq34}
M = \begin{pmatrix}
1&0&0&0\\
0&\lambda_{1}(p)&0&0\\
0&0&\lambda_{2}(p)&0\\
0&0&0&\lambda_{3}(p)
\end{pmatrix},   
\end{equation}
where, $\lambda_{1}(p) =\lambda_{2}(p)=\lambda_{3}(p) = \frac{3}{4}  \alpha p^{2} - \alpha p  - p +1$. The matrix $M$ can be written as 
\begin{equation}\label{eq35}
    M = \begin{pmatrix}
1&\textbf{0}\\
\textbf{0}&\textbf{B}
\end{pmatrix}.
\end{equation}
Here, \textbf{B} is the matrix satisfying condition $|\mathbf{B}|=|M|$.
The absolute value of the determinant of matrix $M$, $|M|$, can be associated with volume of accessible states \cite{lang2012introduction}. It gives the reduction factor for the volume of accessible states, given by the measure of the set of evolved Bloch vectors, with respect to its value at $p = 0$.  We plot $|M(p)|$ as a function of  $p$ for the  evolution of the Bloch vectors under the action of  the general depolarizing  noisy map in Fig. \ref{fig:fig7} for $\alpha = 0$ (no perturbation; left) and  $\alpha = 0.8$ (with perturbation), right. When $\alpha=0$,  the  evolution of the norm
of the map  demonstrates  Markovian  dynamics, as can be discerned from the monotonic decrease in $|M(p)|$. For  $\alpha=0.8$, increase in the value of $|M(p)|$ is observed, for $p$ ranging from  $0.7$ to $1.0$,indicating non-Markovianity. In spherical co-ordinates, it is then straightforward to check that any positive trace-preserving map described by
Eq. (\ref{eq32}) induces the change
\begin{equation}\label{eq36}
    \frac{d V}{dp} = ||\mathbf{B} || \cdot  \frac{dV}{dp}|_{p=0},
\end{equation}
where $||\mathbf{B}||$  decreases monotonically for any positive, linear, and trace preserving map \cite{wolf, lorenzo}. The values of $||\textbf{B}||$ as well as $||M(p)||_{1}$ could be used to identify non-Markovianity, as a similar trend like $|M(p)|$ will be observed for them.
The $\frac{d ||M(p)||_{1}}{dp} >0$ for some $p$, is then indicative of non-Markovian dynamics. Using this, the quantification of non-Markovianity for the depolarizing channels is 
\begin{align} \label{eq37}
    N_{v}=  \int _{\alpha_{-}}^{1.0} \frac{d ||M(p)||_{1}}{dp} dp=\frac{3}{4}\alpha. 
\end{align}
 It is worth mentioning  that  the $\mathcal{N}_{BLP}$ and $\mathcal{N}_{v}$ measures show similarity. 

\subsection{ Non-Markovianity through visualizing the trajectory traced by map in the parameter space}\label{sec:level7}

The diagonalized form of the matrix elements of the intermediate map $\Phi(p,q)$ can be defined by three real parameters $\lambda_{1}(p), \lambda_{2}(p)$, and $\lambda_{3}(p)$, as shown in Eqs. (\ref{eq11}) and (\ref{eq12}). An arbitrary Pauli map is defined as \cite{mary,filippov}
\begin{align}\label{eq38}
\Phi(\rho) &=\frac{1}{2}  \left(\text{tr}(\rho) \mathbb{I} + \sum_{i=1}^{3} \lambda_{i}(p)\text{tr}(\sigma_{i}\rho) \sigma_{i}\right) \nonumber \\
&=
\begin{pmatrix}
1&0&0&0\\
0&\lambda_{1}(p)&0&0\\
0&0&\lambda_{2}(p)&0\\
0&0&0&\lambda_{3}(p)
\end{pmatrix}.
\end{align}
Following \cite{mary,filippov} the map $\Phi(\rho)$ is  positive if $-1 \leq \lambda_{i} \leq 1$ (cube in the parameter space) and the map $\Phi(\rho)$ is CP if $1\pm \lambda_{3}(p) \geq |\lambda_{1}(p) \pm \lambda_{2}(p) |$. The smooth process $\Phi(\rho)$ can then be determined by a continuous trajectory ${\bf{ \lambda}}(p)$ in the parameter space. Such a trajectory provides a pictorial representation of the dynamical map in $\mathbb{R}^{3}$. By analyzing the process trajectory in the parameter space, its (non-)Markovian properties are revealed.  In \cite{filippov},  the dynamical maps were visualized via paths in the parameter space and their divisibility was investigated. The dynamics loses CP divisibility under infinitesimal perturbations. 
Any qubit Pauli map   $\Phi(\rho)$ between finite dimensional spaces can be described by three  real parameters $\left(\lambda_{1}(p), \lambda_{2}(p),\lambda_{3}(p)\right)$ \cite{hall, wudarski, chruscinski,Chruscinski2}. Investigating the process trajectory in the parameter space brings out its divisibility properties. Our goal here is to detect the  non-Markovianity  of the depolarizing map  in terms of its trajectory in the parameter space \cite{ruskai,Bengtsson,filippov}.  

\begin{figure}[htpb] 
    \centering
    \includegraphics[width=9cm]{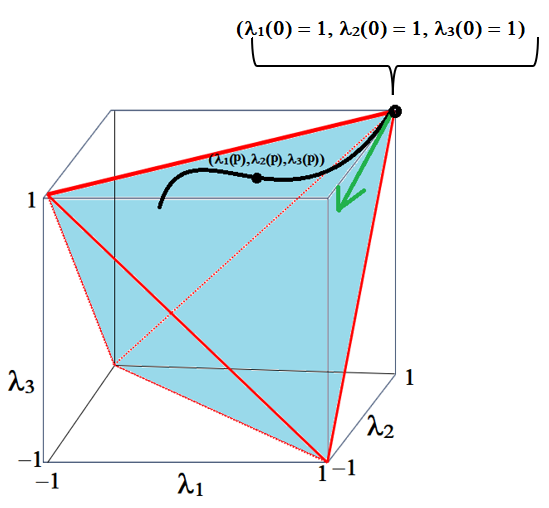}
\caption{ Geometrical presentation of the map where the curve corresponds to  CP maps since it is located inside the tetrahedron. The vector \textbf{A(p)} is drawn from corner (1,1,1) and is pointing inside the tetrahedron, indicating the "CP-divisibility.}
\label{fig:fig8}
\end{figure}

For any linear map, $\Phi(t,0)$, the trajectory $\{ \lambda_{1},  \lambda_{2},   \lambda_{3} \}$ can be an arbitrary curve inside the tetrahedron $1\pm \lambda_{3}\geq |\lambda_{1}\pm \lambda_{2}|$; Fig. \ref{fig:fig8}. The curve starts initially, at $p=0$, from the corner $(\lambda_{1}(0) = 1,\lambda_{2}(0) = 1,\lambda_{3}(0) = 1)$. If the traced trajectory curve stays in the tetrahedral part of the positive octant made from the cube shown in Fig. \ref{fig:fig7}, the corresponding map $\phi_{t}$ is said to be CP. The trajectory of the curve residing in this positive octant pertains to some map $\phi_{t}$ with positive decoherence rates. The curve traversing out of this octant is indicative of a negative decoherence rate and non-CP behavior \cite{filippov}.

\begin{figure*}[htpb]
\centering
\subfloat[Markovian evolution]{\includegraphics[height=6cm]{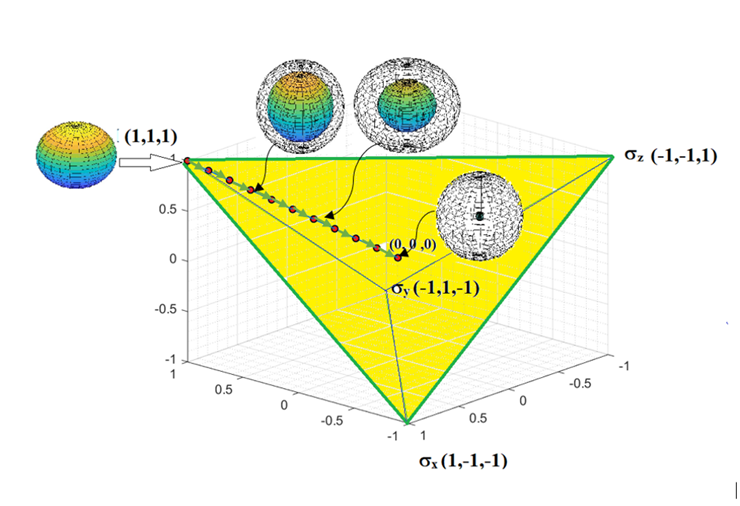}} 
\subfloat[non-Markovian evolution]{\includegraphics[height=6cm]{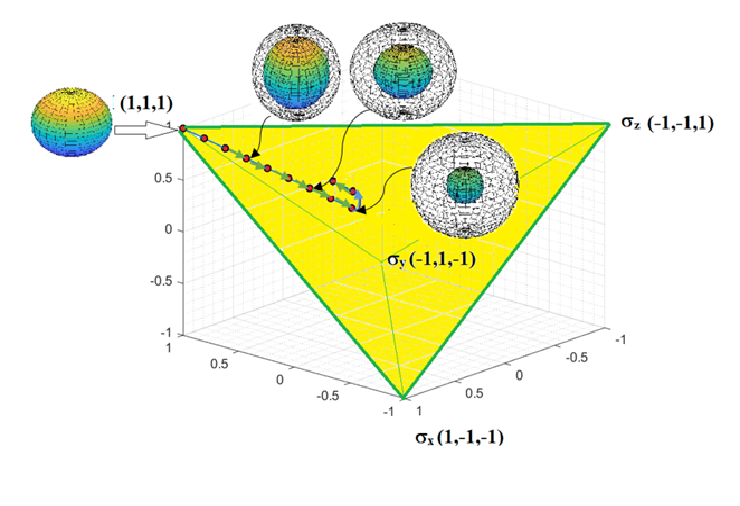}}
\caption{Trajectory-based visualization of dynamics (a) for $\alpha=0$ and  (b) for $\alpha=0.7$.}
\label{fig:fig9}
\end{figure*}

To discover direction during the dynamical progress in the parameter space, we use the following formula \cite{filippov}:
\begin{equation}\label{eq39}
     \dot{\Phi} \circ \Phi^{-1} \left(\rho \right) = \frac{1}{2} \sum_{i=1}^{3} 
     \frac{\dot{\lambda_{i}}(p)}{\lambda_{i}(p)} tr(\rho) \sigma_{i}  
\end{equation}
Accordingly, we define a vector $\bf{A}(p)$ as 
\begin{equation}\label{eq40}
   {\bf{A}}(p) = \left(\frac{\dot{\lambda}_{1}(p)}{\lambda_{1}(p)},\frac{\dot{\lambda}_{2}(p)}{\lambda_{2}(p)},\frac{\dot{\lambda}_{3}(p)}{\lambda_{3}(p)} \right),
\end{equation}
where $\dot{\lambda}_{i}(p) \equiv \frac{d{\lambda}_{i}(p)}{dp}$.  The vector $\bf{A}(p)$ can be established at any value $p$, highlighting the non-Markovian property of the dynamics. The dynamics is CP divisible if and only if the vector $\bf{A}(p)$ drawn from the corner $(1,1,1)$ of the parameter space points inside the tetrahedron in Fig. \ref{fig:fig8}. The scalar products of $\bf{A}(p)$  with vectors $(-1, 1, 1)$, $(1, -1, 1)$, and $(1, 1, -1)$ are all nonpositive,
\begin{subequations}
\begin{align}
     -A_{1}(p) + A_{2}(p) + A_{3} (p) \leq 0 , \label{eq41} \\
      A_{1}(p) - A_{2}(p) + A_{3} (p) \leq 0, \label{eq42} \\
      A_{1}(p) + A_{2}(p) - A_{3} (p) \leq 0 .\label{eq43}
\end{align}
\end{subequations} 
If a vector $\bf{A}(p)$ fails to satisfy Eq. (\ref{eq41})-(\ref{eq43}), it indicates the non-Markovian CP-indivisible nature of the dynamics.
For a general depolarizing map, its trajectory is given by $\lambda(p) = (\lambda_{1}(p), \lambda_{2}(p),\lambda_{3}(p)) = \frac{3}{4} \alpha p^{2} - \alpha p -p +1$. In the case of Markovian dynamics, where $\alpha = 0$ (no perturbation), the expressions become $A_{1}(p) = A_{2}(p) = A_{3}(p) = \left(1-p\right)$, and $\dot{\lambda}_{1}(p)= \dot{\lambda}_{2}(p) = \dot{\lambda}_{3}(p) = -1$. This results in $A_{1}(p) = A_{2}(p) = A_{3}(p) = \frac{-1}{1-p}< 0 $, for $p \in [0,1]$.  The trajectory curve, depicted in Fig. (\ref{fig:fig8}a), is inside the positive octant; the curve traverses a path from $(1,1,1)$ to $(0,0,0)$. Consequently, the map $\Phi(p,0)$ satisfies Eqs. (\ref{eq41})-(\ref{eq43}), indicating that the dynamics is CP divisible. Conversely, when $\alpha=0.7$ (with perturbation), the curve is modified to visualize the absolute value of the eigenvalues. It is then observed that the curve traverses a reverse path for $p > \alpha_{-}(\alpha=0.7)$. This shift is shown in the plot given in Fig. \ref{fig:fig8}(b). The expressions for $A_{1}(p)$, $A_{2}(p)$, and $A_{3}(p)$ become $\left(\frac{42p - 68}{21p^{2} - 68p + 40}\right)$. The inequalities in Eqs. (\ref{eq37})-(\ref{eq39}) are violated for the intermediate map when $0.8 \leq p \leq 1$. Hence, $\Phi(p,0)$ is non-Markovian in this interval, consistent with our findings in the previous section.


\section{Brief overview of depolarizing noise in higher dimensional systems}\label{sec:level4add}

An understanding of non-Markovianity was attempted in previous sections using several dynamical and geometric characterization techniques at the qubit level. Here, a foray is made into non-Markovianity exhibited by higher dimensional systems, evolving through the depolarizing channel, using a dynamical and geometrical characterization method.  \par

We define non-Markovian depolarizing noise in an $N$-level qudit system \cite{dutta2023qudit} using the following Kraus operator

\begin{widetext}
\begin{equation}\label{eq:qutrit_map}
    E_{r,s}= \begin{cases}
    \sqrt{\left(1-\frac{(N^2-1)}{N^2}\alpha p \right)\left(1-\frac{(N^2-1)}{N^2} p\right)}\mathbb{I}_{N} & \text{for $r= 0 \And s=0$};\\
    \sqrt{1+ \alpha\left(1-\frac{(N^2-1)}{N^2}p\right)\frac{p}{N^2}} U_{r,s} & \text{for $0 \leq r,s \leq (N-1)$, $(r,s)\neq( 0,0)$ },
    \end{cases}
\end{equation}
\end{widetext}
where $U_{r,s}$ is the Weyl operator defined as 
\begin{equation}
    U_{r,s}= \sum_{i=0}^{N-1}\omega^{ir}_{N}\ket{i}\bra{i\oplus s}, 0 \leq r,s \leq (N-1),
\end{equation}
where $\omega_{N}=\exp{\left( \frac{2 \pi \iota}{N} \right)}$, and $\oplus$ denotes addition modulo $N$. These Kraus operators are now used in the study of non-Markovianity in higher dimensional space. The N-level depolarizing map provides a complex example with singularities, and non-Markovianity. Understanding the variation in non-Markovianity and singularity structure with increased system dimension is a worthwhile exploration. An attempt is made in this direction for a three-level qutrit system, where the number of Kraus operators given in Eq. (\ref{eq:qutrit_map}) is nine. Extensions to higher dimensions can be made in a similar fashion. 

\subsection{non-Markovianity for qutrit system (N=3)}
We provide a perspective to non-Markovian effects in a qutrit system undergoing depolarizing noise. We present the dynamical characterization of these effects using the HCLA measure, which relies on negative decoherence rates. In addition the geometrical visualization of non-Markovian effects is also attempted using volume change in the physical states of the system. 

\subsubsection{HCLA Measure}
\begin{figure}[htpb]
    \centering
    \includegraphics[width=9cm]{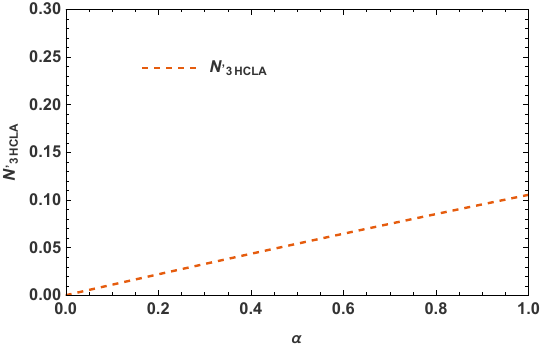}    
\caption{A plot of HCLA measure for qutrit system.}
\label{fig:qutrit_HCLA}
\end{figure}

For an $N$-level qudit system, $k(p)= p + \alpha p -\frac{(N^2-1)}{N^2}\alpha p^2$, [see Eq. (\ref{eq6})]. In the case of qutrit, it becomes $k(p)= p + \alpha p -\frac{8}{9}\alpha p^2$. Following the normalization procedure described in Sec. \ref{sec:HCLA}, we obtain normalized rates. The integration of normalized rates from the point $\alpha_{-}$, obtained from solving the quadratic equation $9p +9 \alpha p - 8 \alpha p^2 -9=0$, to $1$ results in 
\begin{align}
    \label{eq:qutrit_HCLA}
    \mathcal{N^{'}_{\text{3}HCLA}} &= \int_{\alpha_{-}}^{1}\tilde{\gamma} 
    (p) dp \nonumber \\
    & =  \Biggl\{ \ln \left(p \right) + \ln \left(9 + 9\alpha - 8 p \alpha  \right) \Biggr\}^{p=1}_{p=\alpha_{-}}.\nonumber\\
\end{align}
Figure \ref{fig:qutrit_HCLA} depicts the variation in non-Markovianity as a function of parameter $\alpha$. The increase in $\alpha$ results in higher non-Markovianity. Comparing the non-Markovianity in qubit and qutrit systems, one can observe a decrease with increase in dimension. The point $\alpha_{-}$ shifts to a higher value with increasing levels of the system, shrinking the region of negative decay rates. This observation further points to the vanishing of the characteristic singularity structure for systems with $N\geq 5$ for $p,\alpha$ ranging from $0$ to $1$.

\begin{figure}[htpb]
    \centering
    \includegraphics[width=9cm]{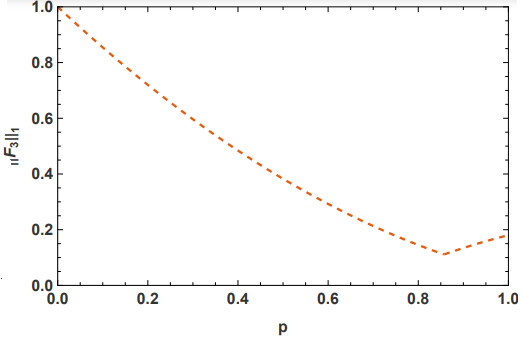}    
\caption{The variation in $||F_{3}||_{1}$ for $\alpha=0.7$. }
\label{fig:fig11}
\end{figure}

\subsubsection{Variation in volume of physical states}
The geometric depiction of non-Markovian effects can be visualized using the variation in the volume of physical states, as in the previous Sec. \ref{sec:level6}. We extend this idea to the $N$-level system and also present the analysis of qutrit system (N=3). 
The generalized Gell-Mann matrix \cite{bertlmann2008bloch} satisfies the properties

\begin{equation}
    G_{0}=\frac{1}{\sqrt{d}}\mathbb{I}_{d}; \quad G_{m}= G_{m}^{\dagger} ; \quad \text{tr}[G_{m}G_{n}]=2\delta_{mn},
\end{equation}
and can be used. For N=3, this results in nine basis operators given as 
\begin{equation}
    G= \{\frac{1}{\sqrt{3}}\mathbb{I}_3,\frac{1}{\sqrt{2}} G_{m} \},
\end{equation}
where $G_{m=1,2,..8}$ represents Gell-Mann matrices. Using these basis operators and the depolarizing channel map constructed using the operators defined in Eq. (\ref{eq:qutrit_map}), we construct the $F_{3}$ matrix \cite{andersson2007finding,sabale2024facets}, whose elements are 
\begin{equation}
    F_{kl}:=\frac{1}{3^2}\text{tr}[G_{k}\mathcal{E}_{p}(G_{l})].
    \label{eq:F_mat}
\end{equation}
The non-monotonic variation in $||F_{3}||_{1}$ is characteristic of non-Markovian dynamics and is depicted in Fig. \ref{fig:fig11} for the qutrit system, by the nonmonotonic behavior. Similar results can be obtained for higher dimensional systems.

\begin{figure*}[htpb]
\centering
\subfloat[multi-qubit map]{\includegraphics[height=7cm]{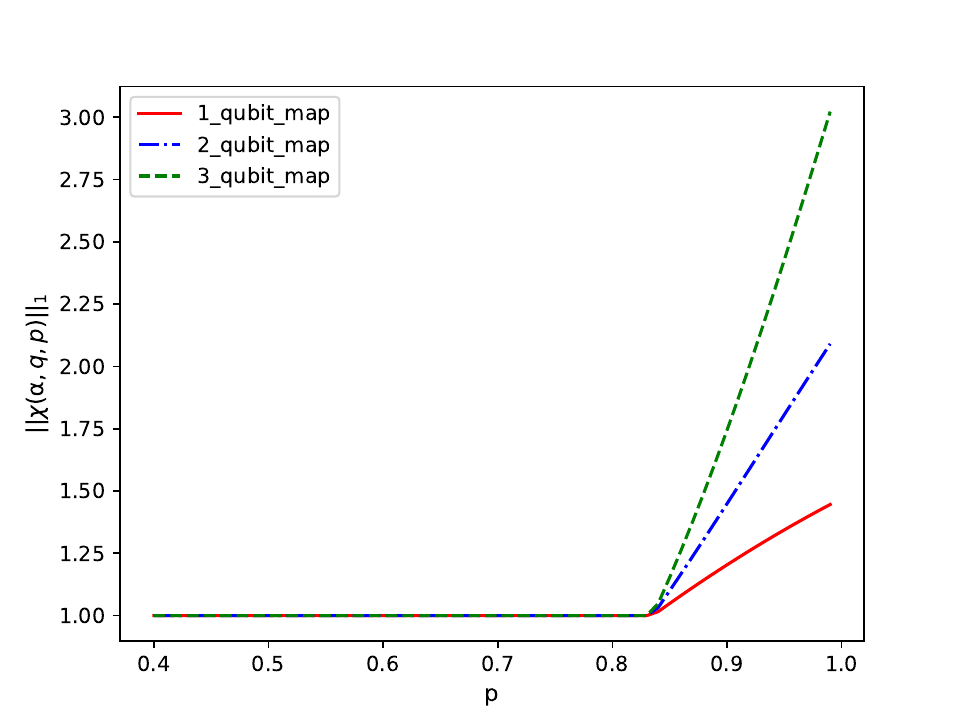}} 
\subfloat[multi-levle map]{\includegraphics[height=7cm]{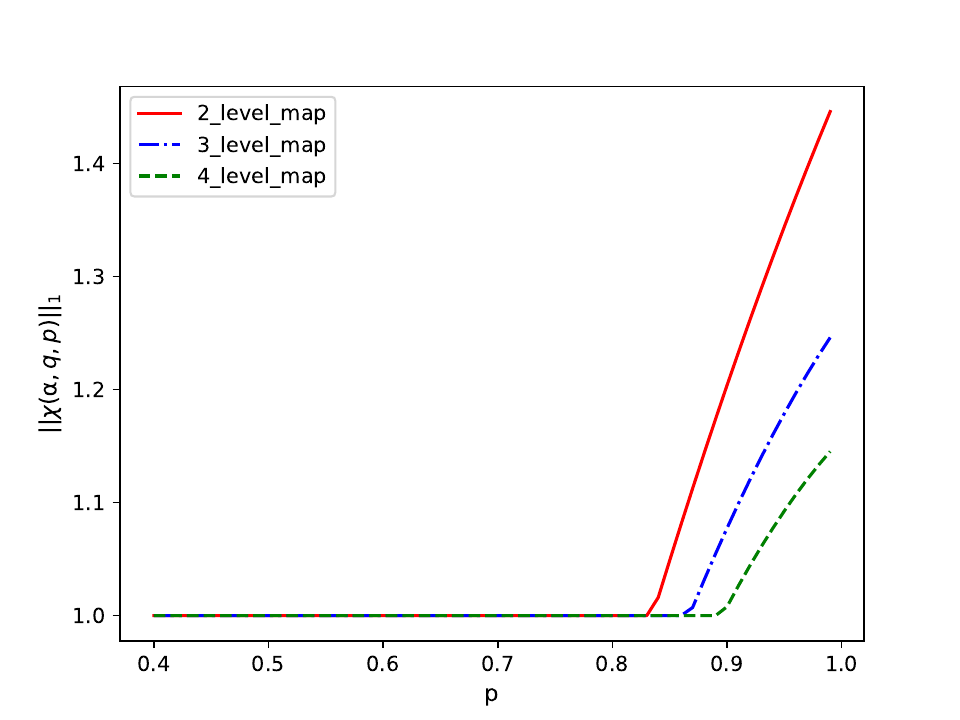}}
\caption{ The variation in $||\chi(\alpha, q, p)||_{1}$ with $\alpha=0.9, q=0.4$, for a (a) multiqubit map, and (b) multi-level map, highlights the impact on the singularity structure, which is invariant for a multiqubit map and changes for a multi-level map affecting non-Markovianity.}
\label{fig:fig12}
\end{figure*}

 \section{Investigating the non-Markovianity of depolarizing noise in multi-qubit Systems}\label{sec:level4new}

In the previous section, we extended our investigation to encompass higher-dimensional systems, examining their non-Markovian behavior as they evolve through the depolarizing channel using both dynamical and geometric characterization approaches. We now turn our attention to the non-Markovianity observed in multiqubit systems.
We examine the scenario in which qubits are exposed to independent depolarizing channels that can be adjusted to display both Markovian and non-Markovian dynamics \cite{PhysRevA.93.032135}. Then, the dynamical evolution of the multi-qubit systems will be obtained by utilizing the Kraus operators constructed from the tensor product of single qubit Kraus operators [Eq. \ref{eq5}]. 
Following the previously used method in Sec. \ref{secIIA} and Appendix \ref{sec:appendix}, one can construct the Choi matrices $\chi(\alpha, q, p)$ of multiqubit maps. \par
\begin{figure}[htpb]
    \centering
    \includegraphics[width=9cm]{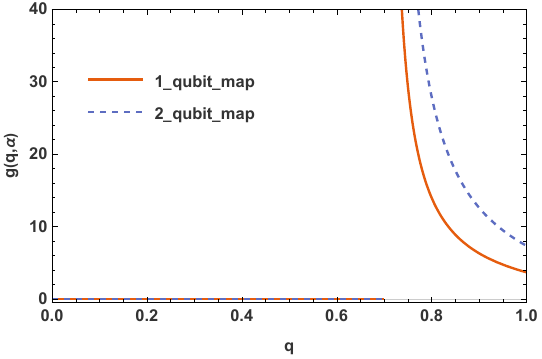}    
\caption{The variation in $g(q,\alpha )$ for $\alpha=0.9$. }
\label{fig:fig13}
\end{figure}

The variation in  trace norm of $\chi(\alpha, q, p)$ is useful to probe the singularity structure and NCP nature of higher-qubit and higher-level maps. The numerical variation in the trace norm of $\chi(\alpha, q, p)$ is shown in Fig. \ref{fig:fig12}(a), depicting the invariant singularity structure due to the same $\alpha_{-}$ in the case of one-, two-, and three-qubit maps. Similarly, Fig. \ref{fig:fig12}(b) reveals the increase in value of $\alpha_{-}$, indicating the possibility of the vanishing singularity structure and non-Markovianity with increasing levels of quantum system. Further, from the analysis of the two-qubit channel, using the right derivative of the trace norm of $\chi(\alpha, q, q+\epsilon)$  results in the function \cite{rivas2} 
\begin{equation}
  g(q,\alpha )=\lim_{x\to\ 0^{+}}\frac{||\chi(\alpha, q, q+\epsilon)||_{1}-1}{\epsilon}.
\end{equation}
This reveals enhanced non-Markovian features in the two-qubit map compared to the single-qubit map. The variation of $g(q,\alpha )$ is depicted in Fig. \ref{fig:fig13}. The two-qubit map reaches higher values of the function $g(q,\alpha )$ compared to the single-qubit map, indicating enhanced non-Markovianity.

\section{Conclusion}\label{sec:level5}
The notion of non-Markovianity in completely positive depolarizing noise is explored using various dynamical and geometrical techniques. The introduced non-Markovianity parameter ($\alpha$) of the depolarizing map, is the cause of perturbations and non-Markovianity. It contributes to breaking the CP divisibility of the intermediate map, assuring non-Markovianity. The non-Markovian behavior of the depolarizing map could be identified by using appropriate witness operators. The intermediate evolution admits information backflow, showing non-monotonic variation in quantum memory, which could be justified using witness $X_{\chi(\alpha, q, p)}$. The witnessed non-Markovianity is quantified using two techniques: (a) BLP measure based on the distinguishability of states, and (b) HCLA measure based on the negativity of the decoherence rates. We further study geometric visualization techniques to gain a comprehensive understanding of non-Markovianity. It is observed that non-Markovianity contributes to enhancing the accessible volume of states during evolution. Furthermore, the trajectory traced during evolution in parameter space presents the loss of the CP divisibility under perturbations caused by non-zero $\alpha$. The extension of this analysis to higher dimensional systems is made and illustrated for a qutrit system. The non-Markovianity and singularity structure are observed to decreases with increase in system size. Finally, this analysis is extended to multi-qubit systems, where it is observed that non-Markovianity increases with the number of qubits. The singularity structure remains invariant in multiqubit maps; however, it is affected and ultimately lost in higher-dimensional (qudit) maps. 


\appendix

\section{ Constructing the   Choi matrix  $\chi (\alpha, q,p) $ for  the intermediate map \label{sec:appendix} }
In this appendix, we present the detailed mathematical  construction  of the Choi matrix for  the intermediate map (propagator). We begin by constructing $U_{2\leftrightarrows 3}[\Phi(p,q) \otimes  \mathbb{I}_{N^2}]U_{2\leftrightarrows 3}$ where $U_{2\leftrightarrows 3}$, for $N$-dimensional system, is the commutation
(or ``swap") matrix \cite{horn1994topics,magnus} between the ``second" and the ``third" subspaces \cite{rivas2}. Second, we
apply $U_{2\leftrightarrows 3}[\Phi(p,q) \otimes  \mathbb{I}_{N^{2}}]U_{2\leftrightarrows 3}$ on $ \rm{vec}(\ket{\Psi}\bra{\Psi})$, where $\ket{\Psi}= \sum_{i=0}^{N-1} \frac{1}{\sqrt{N}}\ket{ii}$.  $U_{2\leftrightarrows 3}$ is written as $\mathbb{I}_{N} \otimes  U_{P} \otimes  \mathbb{I}_{N}$, where $U_{P}$ is a permutation matrix satisfying the property $U_{P}(A\otimes B)U_{P}= B \otimes A$, and is defined as
\begin{equation}
U_{P} = \sum_{k=0}^{N-1}\sum_{l=0}^{N-1} \ket{k}\bra{l}\otimes \ket{l}\bra{k}.    
\end{equation}
We assume that $\phi(p,q) \otimes  \mathbb{I}_{N^2}$ acts as  a tensor product of four
spaces with the same dimension, $\mathcal{H}_{1}\otimes\mathcal{H}_{2} \otimes \mathcal{H}_{3} \otimes\mathcal{H}_{4}$. Then $U_{2\leftrightarrows 3}$ denotes the permutation matrix  interchanging the second and third subspace, $i.e.$, $U_{2\leftrightarrows 3} = \mathbb{I}_{N} \otimes  U_{P} \otimes  \mathbb{I}_{N}$ . Lastly, we write the result as a matrix, $i.e.$, ``devectorize" to construct the Choi matrix  of the intermediate map as follows
\begin{equation}
    \chi(\alpha, q,p)  = \left[ \Phi(p,q) \otimes \mathbb{I}  \right]\ket{\Psi} \bra{\Psi}.     
\end{equation}
Thus, we obtain the matrix in Eq. (\ref{eq11}) for the case of $N=2$.


\bibliography{main}

\end{document}